\documentclass[a4paper,fleqn,usenatbib]{mnras}

\usepackage[T1]{fontenc}
\usepackage{ae,aecompl}

\usepackage{graphicx}	% Including figure files
\usepackage{amssymb}	% Extra maths symbols
\usepackage[usenames,dvipsnames]{color} %Text color

\newcommand{\alm}{a_{\ell m}}

\newcommand{\Cee}{\mathcal{C}}
\newcommand{\Cl}{\Cee_\ell}

\newcommand{\Ccorr}{\Cee} % C(theta) -> \Ccorr(\theta)

\newcommand{\ellmin}{\ell_{\rmn{min}}}
\newcommand{\ellmax}{\ell_{\rmn{max}}}

\newcommand{\Nside}{\textit{Nside}}

\newcommand{\muK}{\,\rmn{\umu K}}
\newcommand*{\code}[1]{\textsc{#1}}

\newcommand{\healpix}{\code{healpix}}
\newcommand{\camb}{\code{Camb}}

\newcommand*{\satellite}[1]{\textit{#1}}

\newcommand{\COBEDMR}{\satellite{COBE-DMR}}
\newcommand{\WMAP}{\satellite{WMAP}}
\newcommand{\Planck}{\satellite{Planck}}

\newcommand*{\Planckmap}[1]{\texttt{#1}}
\newcommand{\smica}{\Planckmap{SMICA}}
\newcommand{\nilc}{\Planckmap{NILC}}
\newcommand{\sevem}{\Planckmap{SEVEM}}
\newcommand{\commander}{\Planckmap{Commander}}
\newcommand{\common}{\Planckmap{Common}} % Common mask

\newcommand{\LCDM}{$\Lambda$CDM}

\renewcommand*{\vec}[1]{\bmath{#1}}
\newcommand*{\unitvec}[1]{\vec{\hat{#1}}}

\newcommand{\Shalf}{S_{1/2}}

\newcommand{\eISW}{eISW}
\newcommand{\lISW}{\ifmmode\ell\rmn{ISW}\else$\ell$ISW\fi}

\newcommand{\early}{e}
\newcommand{\late}{l}

\newcommand{\sI}{\textit{unconditioned-$\Lambda$CDM}}
\newcommand{\sII}{\textit{\smica-conditioned}}
\newcommand{\sIII}{\textit{low-$\Shalf$-\LCDM}}
\newcommand{\sIV}{\textit{low-$\Shalf(\early\early,\early\early)$-\LCDM}}

\def\gsim{\mathrel{\rlap{\lower4pt\hbox{\hskip1pt$\sim$}}
    \raise1pt\hbox{$>$}}}

\newcommand{\iimag}{\rmn{i}}
\newcommand{\dderiv}{\rmn{d}}

\title[The ISW effect and absence of $\Ccorr^{TT}(\theta \gsim 60^\circ)$]{The ISW effect and the lack of large-angle CMB temperature correlations}

\author[C.J. Copi, M. O'Dwyer and G.D. Starkman]
{
Craig J. Copi$^{1}$\thanks{E-mail: cjc5@case.edu},
M\'arcio O'Dwyer$^{1,2}$\thanks{E-mail: marcio.odwyer@case.edu},
Glenn D. Starkman$^{1,3}$\thanks{E-mail: glenn.starkman@case.edu}
\\
$^{1}$CERCA/Department of Physics/ISO, Case Western Reserve University,
 Cleveland, OH 44106-7079, USA\\
$^{2}$The Capes Foundation, Ministry of Education of Brazil, Bras\'ilia DF 70359-970, Brazil\\
$^{3}$Observatorio Nacional, Rio de Janeiro, RJ 20921-400, Brazil\\
}

\date{Accepted XXX. Received YYY; in original form ZZZ}

\pubyear{2016}

\begin{document}
\label{firstpage}
\pagerange{\pageref{firstpage}--\pageref{lastpage}}
\maketitle

% Abstract of the paper
\begin{abstract}
  It is by now well established that the magnitude of the two-point
  angular-correlation function of the cosmic microwave background
  temperature anisotropies is anomalously low for angular separations
  greater than about $60$ degrees.  Physics explanations of this anomaly
  typically focus on the properties of the Universe at the surface of last
  scattering, relying on the fact that large-angle temperature fluctuations
  are dominated by the Sachs-Wolfe effect (SW).  However, these
  fluctuations also receive important contributions from the integrated
  Sachs-Wolfe effect (ISW) at both early (\eISW) and late (\lISW) times.
  Here we study the correlations in those large-angle temperature
  fluctuations and their relative contributions to $\Shalf$ -- the standard
  measure of the correlations on large angular scales.  We find that 
  in the best-fitting \LCDM\ cosmology, 
  while the auto-correlation of the early contributions (SW plus \eISW)
  dominates $\Shalf$, 
  there are also significant contributions originating from
  cross-terms between the early and late contributions.  In particular,
  realizations of \LCDM\ with low $\Shalf$ are typically produced from a
  combination of somewhat low pure-early correlations and accidental
  cancellations among early-late correlations.  We also find that if the
  pure \lISW\ auto-correlations were the only contribution to $\Shalf$ in
  \LCDM, then the $p$-value of the observed cut-sky $\Shalf$ would be
  unremarkable.  This suggests that physical mechanisms
  operating only at or near the last scattering surface could explain the
  observed lack of large-angle correlations, though this is not the typical
  resolution within \LCDM.
\end{abstract}

% Select between one and six entries from the list of approved keywords.
% Don't make up new ones.

%%%%%%%%%%%%%%%%%%%%%%%%%%%%%%%%%%%%%%%%%%%%%%%%%%

%%%%%%%%%%%%%%%%% BODY OF PAPER %%%%%%%%%%%%%%%%%%

\section{Introduction}

The temperature ($T$) and $E$-mode polarization fluctuations of the Cosmic
Microwave Background (CMB) are widely regarded as one of the great
successes of the standard cosmological model -- inflationary Lambda Cold
Dark Matter (\LCDM)\@.  Even so, some anomalies, particularly on large
angular scales, have been identified 
\citep[for reviews of anomalies see][]{2010AdAst2010E..92C,2011ApJS..192...17B,2015arXiv150607135P,2015arXiv151007929S}. 
In this paper we focus
on the first identified anomaly -- the absence of two-point angular
correlation in the CMB temperature maps for angular separations above
approximately $60$ degrees.  While this was originally noticed in the
\COBEDMR\ data \citep{1996ApJ...464L..25H}, it was first quantified by
the \WMAP\ team in their first-year data release  \citep{2003ApJS..148..175S} through a statistic, $\Shalf$, which we
recall below.  This $\Shalf$ statistic has been found to be anomalously low
on the full reconstructed CMB sky and even more so on  the cleanest part of the sky in all full-sky maps since the
first-year \WMAP\ Independent Linear Combination (\Planckmap{ILC}) map and
including all CMB-dominated single-waveband maps 
\citep{2006MNRAS.367...79C,2007PhRvD..75b3507C,Copi:2008hw,2013arXiv1310.3831C}.

The lack of large-angle correlation in the temperature two-point angular
correlation function could be a statistical fluke: our Universe could be a
rare realization of the \LCDM\ ensemble.  
Though we cannot probe other universes from the ensemble,
this `fluke hypothesis' can be tested \citep{2007astro.ph..2723H,2008PhRvD..77f3008D,2013MNRAS.434.3590C,2014MNRAS.442.2392Y,2015PhRvD..91l3504Y}. 
Alternatively, the observed  lack of large-angle correlation may be due to
new physics, the simplest ideas involving modifications at the last
scattering surface (LSS) of the CMB\@.

The dominant contribution to low-$\ell$ $T$ power is the Sachs-Wolfe (SW)
effect on the LSS -- the effect on photon energy due to the difference
between the Newtonian potential on the LSS and the potential at the
observer.  Much of the thought (so far unsuccessful) into how to explain
the low-$\Shalf$ anomaly has therefore concentrated on eliminating
large-angle correlations in the gravitational potential at last scattering.
However, the integrated Sachs-Wolfe (ISW) effect -- the result of photon
propagation through a time-dependent potential -- also contributes
significantly at low-$\ell$ and large angles.  This ISW effect itself can
be divided into two major contributions -- early-time (\eISW), i.e. at high
redshift, and late-time (\lISW), i.e. at low redshift.

Presumably, any mechanism that de-correlated the gravitational potential on
the LSS would eliminate correlations in both the SW and the \eISW\@. The
same is not true for the \lISW\@.  The obvious question then becomes -- is
it sufficient to eliminate correlations on or near the LSS in order to
explain the smallness of $\Shalf$, or must whatever physics is at work also
affect the \lISW -- either its auto-correlation or its cross-correlation
with early-time physics?

We address this question by studying how the rare small-$\Shalf$ skies
emerge by chance within \LCDM\@.  We do this by separately analysing
contributions from early physics, late physics, and their correlations.  We
find that the typical means by which \LCDM\ produces a rare realization
with a low $\Shalf$ is by somewhat lowering the contribution from
pure-early correlations and further reducing it through accidental
cancellations due to early-late cross-correlations.
Alternatively, we find evidence to suggest that postulating new physics at or
near the LSS also has the potential to successfully explain the observed
low value of $\Shalf$.

\section{Formalism}
\label{sec:formalism}

Here we review the standard methods of describing temperature fluctuations
and define the notation employed. The temperature two-point angular
correlation function,
\begin{equation}
  \Ccorr(\theta) \equiv \overline{T(\unitvec e_1) T(\unitvec e_2)}, \qquad
  \unitvec e_1 \cdot \unitvec e_2 = \cos \theta, 
  \label{eq:Ctheta-full-sky}
\end{equation}
is defined as the average over the sky (or some portion thereof) 
of the product between the temperature fluctuations $T(\unitvec e)$ in two directions 
separated by an angle $\theta$. 
On a full sky, it contains the same information as the more familiar (in CMB physics) 
angular power spectrum, $\Cl$, since 
\begin{equation}
\label{eqn:Ccorr}
  \Ccorr(\theta) = \sum_\ell \frac{2\ell+1}{4\upi} \Cl P_\ell(\cos\theta),
  \label{eq:Ctheta-full-sky-expansion}
\end{equation}
where $P_\ell$ is the Legendre polynomial of order $\ell$. 
Here $\Cl$ is given by
\begin{equation}
  \Cl \equiv \frac1{2\ell+1} \sum_{m}|\alm|^2.
  \label{eq:Cl-full-sky}
\end{equation}
and the $\alm$ are the coefficients of a spherical-harmonic expansion of
the CMB temperature fluctuations
\begin{equation}
  T(\unitvec e) \equiv \sum_{\ell m} \alm Y_{\ell m}(\unitvec e).
\end{equation}
Nominally, the lower limit of the sum in
(\ref{eq:Ctheta-full-sky-expansion}) should be $\ellmin=1$; however, since
any intrinsic dipole is presumed to be overwhelmed by a much larger Doppler
dipole from which we are currently unable to separate it, all maps are
monopole and dipole-subtracted, so $\ellmin=2$.

To quantify the observed lack of correlation above $60$ degrees in their
first year data, the \WMAP\ team \citep{2003ApJS..148..175S} introduced the
$\Shalf$ statistic,
\begin{equation}
  \label{eq:Shalf}
	S_{1/2} \equiv \int_{-1}^{1/2} [\Ccorr(\theta)]^2\,\dderiv(\cos\theta).
\end{equation}
This integral can more conveniently be rewritten as the sum
\begin{equation}
  \Shalf = \sum_{\ell \ell'} \Cl I_{\ell\ell'} \Cee_{\ell'},
  \label{eq:S12-sum}
\end{equation}
where $I_{\ell\ell'}$ are elements of a known matrix \citep{Copi:2008hw}.
In the \WMAP\ year one data, the $p$-value of $\Shalf$ in an ensemble of
realizations of the best-fitting \WMAP\  year one \LCDM\ model was
approximately $0.15$ per cent.  
Since then, through all
the \WMAP\ and \Planck\ data releases, with the various full-sky reconstruction
algorithms and associated masks, that $p$-value has varied from about $0.03$
per cent to $0.3$ per cent for masks with this same cut-sky fraction.
It is slightly higher (about $0.6$ per cent) for the \Planck\ Release-2
full-sky maps with the \common\ (UT78) mask,  which only
removes $22$ per cent of the sky.

\section{Separating early-time and late-time effects}
\label{sec:early-time}

As remarked above, the CMB temperature power spectrum at low $\ell$, or
alternatively $\Ccorr(\theta)$ at large angles, is dominated by two
contributions, the Sachs-Wolfe effect and the integrated Sachs-Wolfe effect.
As also noted above, the early-time effects
(SW and \eISW) occur at high redshift, whereas the late-time effect (\lISW)
occurs at low redshift.  These effects are thus well-separated in redshift,
and we can think of these contributions to $\Ccorr(\theta>60\degr)$ as
being divided into early-time and late-time physics.  With this in mind, we
can analyse early-time and late-time effects separately to determine how
they combine to result in the small $\Shalf$ observed.

To separate early-time and late-time contributions we write the total
spherical-harmonic decomposition coefficients as a sum of their early and
late contributions,
\begin{equation}
  \alm = \alm^\early + \alm^\late,
\end{equation}
where `$\early$' stands for early and `$\late$' for late. This separation
corresponds to
\begin{equation}
  \Cl = \Cl^{\early\early} + \Cl^{\late\late} + 2\Cl^{\early\late},
  \label{eq:Cl-breakdown}
\end{equation} 
when written in terms of the power spectra.
Here the superscripts represent two-point auto and cross-correlations 
between early-time and late-time anisotropies.
With this separation the $\Shalf$ sum~(\ref{eq:S12-sum})
can therefore be decomposed into six different
contributions denoted by $S_{1/2} (X,Y)$ with
\begin{equation}
  \Shalf(X,Y) \equiv \sum_{\ell \ell'} \Cl^X I_{\ell\ell'} \Cee_{\ell'}^Y
  \label{eq:S12-XY}
\end{equation}
and $X,Y\in\{\early\early,\late\late,\early\late\}$.  The total $\Shalf$ is the
sum over all of these
contributions.

In \LCDM, the early-time and late-time contributions to $\alm$, the
$\alm^\early$ and $\alm^\late$, are correlated Gaussian random variables
with known correlations. Given values for the $\alm$, either as
realizations of the theory or extracted from cleaned full-sky maps, the
known theoretical correlations can be employed to generate constrained
realizations of pairs of early and late skies from the \LCDM\ ensemble.
(See Appendix \ref{ap:alm} for details.)  From each realization, all the
$\Shalf(X,Y)$ components can be calculated.  To understand how the six
$\Shalf(X,Y)$ components work together to reduce correlations on large
scales (resulting in a low $\Shalf$) we generated realizations of early-sky
and late-sky pairs imposing various constraints on the \LCDM\ ensemble, as
enumerated below.

\subsection{Parameters}

To generate realizations some choices must be made.  Noting that the effects
of interest occur on large angular scales, high resolution maps (or large
multipole moments) are not required to capture the relevant information.
For this reason, all calculations have been performed at a
\healpix\footnote{See \url{http://healpix.sourceforge.net} for more
  information.} resolution of $\Nside=64$ and with $\ellmax=191$.  Further,
the best-fitting \LCDM\ model parameters from \Planck\ ($TT$ + lowP +
lensing) \citep{2015arXiv150201589P} were used along with the January 2015
version of the \camb\footnote{See \url{http://camb.info/} for information
  and access to the code.} code \citep{2000ApJ...538..473L}  to generate
the theory $\Cl$ along with the \lISW\ only contribution ($z<30$),
$\Cl^{\late\late}$, and the early effects (SW+\eISW),
$C_\ell^{\early\early}$.  The cross-correlation, $\Cl^{\early\late}$, is
easily deduced from~(\ref{eq:Cl-breakdown}).

The mask and cleaned maps from \Planck\ are provided at high resolution.
These were degraded to $\Nside=64$ using the method employed in the
\Planck\ Release-2 analyses \citep{2015arXiv150607135P}.  Briefly, the
$\alm$ were extracted from the high resolution maps, scaled by the ratio of
window functions between the new low resolution and original high
resolution, and finally synthesized into a low resolution map. For
degrading masks the additional step of setting all pixels with values less
than $0.9$ to zero, and all others to one, is performed.

Unless otherwise stated, all $\Shalf$ values are calculated from skies
masked by the \Planck\ Release-2 \common\ mask (degraded as described
above) and with the monopole and dipole removed after masking.  This mask
has $f_{\rmn{sky}}=0.78$, slightly larger than used in previous studies
\citep{2013arXiv1310.3831C}. Though this shifts the calculated cut-sky
$\Shalf$ to a larger value, and thus to a larger $p$-value, here we are
only interested in the relative contributions due to early-time and
late-time correlations, not the absolute value.

\subsection{Realizations}
\label{sec:Realizations}

Our analyses are based on sets of realizations generated from different
choices of the total $\alm$.  From these total $\alm$, pairs of early-time,
$\alm^\early$, and late-time, $\alm^\late$, contributions were generated as
discussed in App.~\ref{ap:alm}. The sets of realizations are as follows.

\begin{enumerate}
  \item A total of $200\,000$ realizations of $\alm$ drawn from the
    best-fitting \LCDM\ model were generated.  For each realization, a
    single pair of the $\alm^\early$ and $\alm^\late$ were subsequently
    generated.  This set of realizations represents the expected
    distribution from \LCDM\ and predominantly contains skies with large
    $\Shalf$.  We refer to this set as the \sI\ realizations.
    
  \item The $\alm$ were extracted from each of the cleaned, full-sky
    \Planck\ Release-2 maps (\smica, \commander, \sevem, and \nilc).  A set
    of realizations of pairs of $\alm^\early$ and $\alm^\late$ were
    generated for each map.  Since each of the four Release-2 maps produced
    nearly identical results, only those from the \smica\ map will be
    discussed in detail.  This \smica\ set again contains $200\,000$
    realizations.  We refer to this set as the \sII\ realizations.
    
  \item Ten realizations of $\alm$ drawn from the best-fitting \LCDM\ model
    constrained to have $\Shalf \approx \Shalf(\smica)$ were generated.
    For each of these ten low-$\Shalf$ realizations, $10\,000$ pairs of
    $\alm^\early$ and $\alm^\late$ were subsequently generated.  This set
    of realizations represents \LCDM\ skies with large-angle properties
    similar to those found in the \smica\ map.  It provides a check of
    whether the \sII\ set was typical or atypical among low $\Shalf$
    \LCDM\ realizations.  We refer to these sets collectively as the
    \sIII\ realizations.
    
  \item Subsets from the \sI\ realizations were selected with
    $\Shalf(\early\early,\early\early)\leq\Shalf^{\rmn{max}}(\early\early,\early\early)$,
    for a variety of values of
    $\Shalf^{\rmn{max}}(\early\early,\early\early)$.  In addition, in order
    to probe very small values of $\Shalf(\early\early,\early\early)$, the
    \sI\ realizations were augmented by creating additional realizations
    with $\Shalf(\early\early,\early\early)$ as low as $500\muK^4$.  This
    set of realizations allowed us to probe the effect of suppressed
    early-time correlations in the context of \LCDM\@.  Note that this does
    not represent a physical model for the suppression of early-time
    correlations.  No physics has been put in to cause the suppression.  It
    represents the effect of the \lISW\ on $\Shalf$ in the (near) absence
    of early-time contributions.  We refer to these sets collectively as
    the \sIV\ realizations.
\end{enumerate}

\begin{figure*}
  \includegraphics[width=\columnwidth]{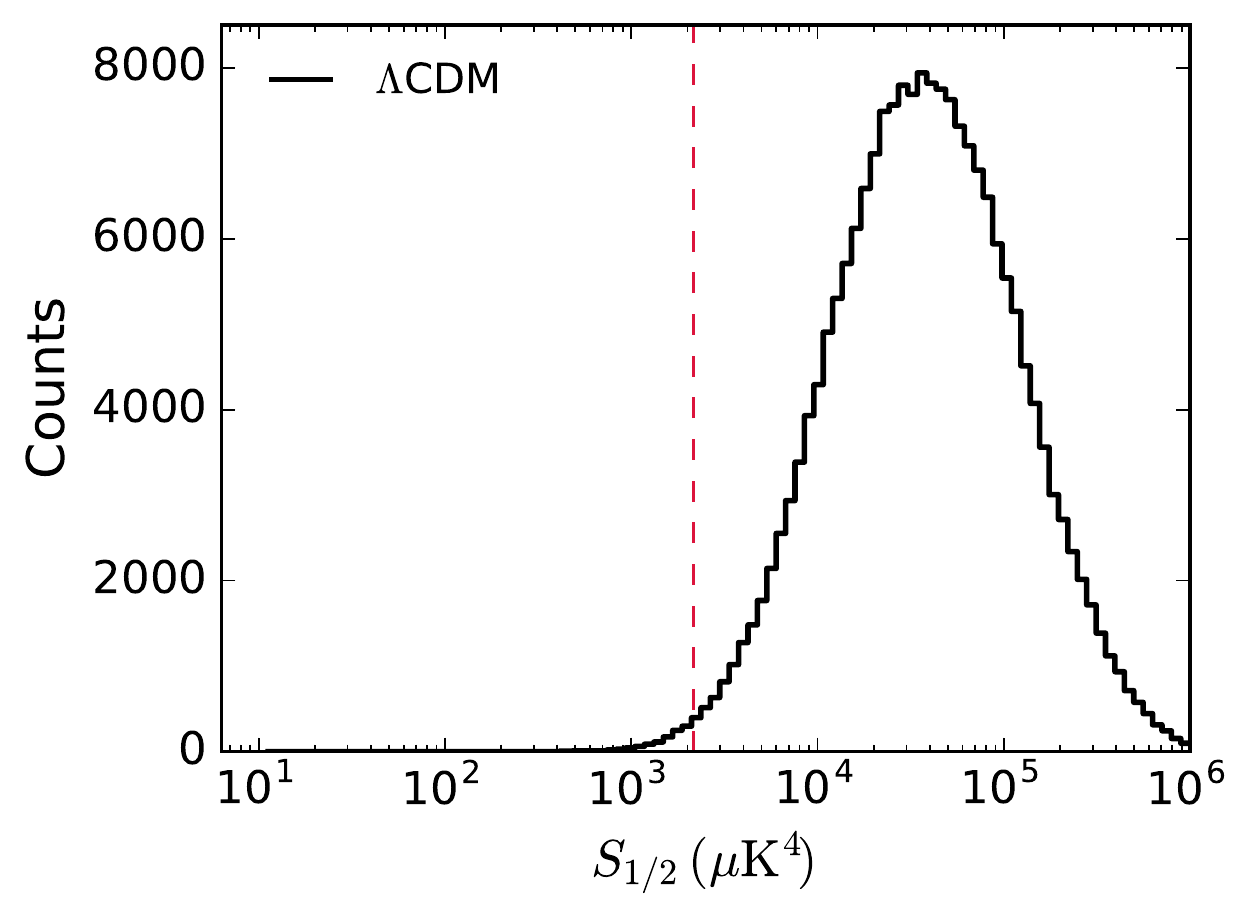}
  \includegraphics[width=\columnwidth]{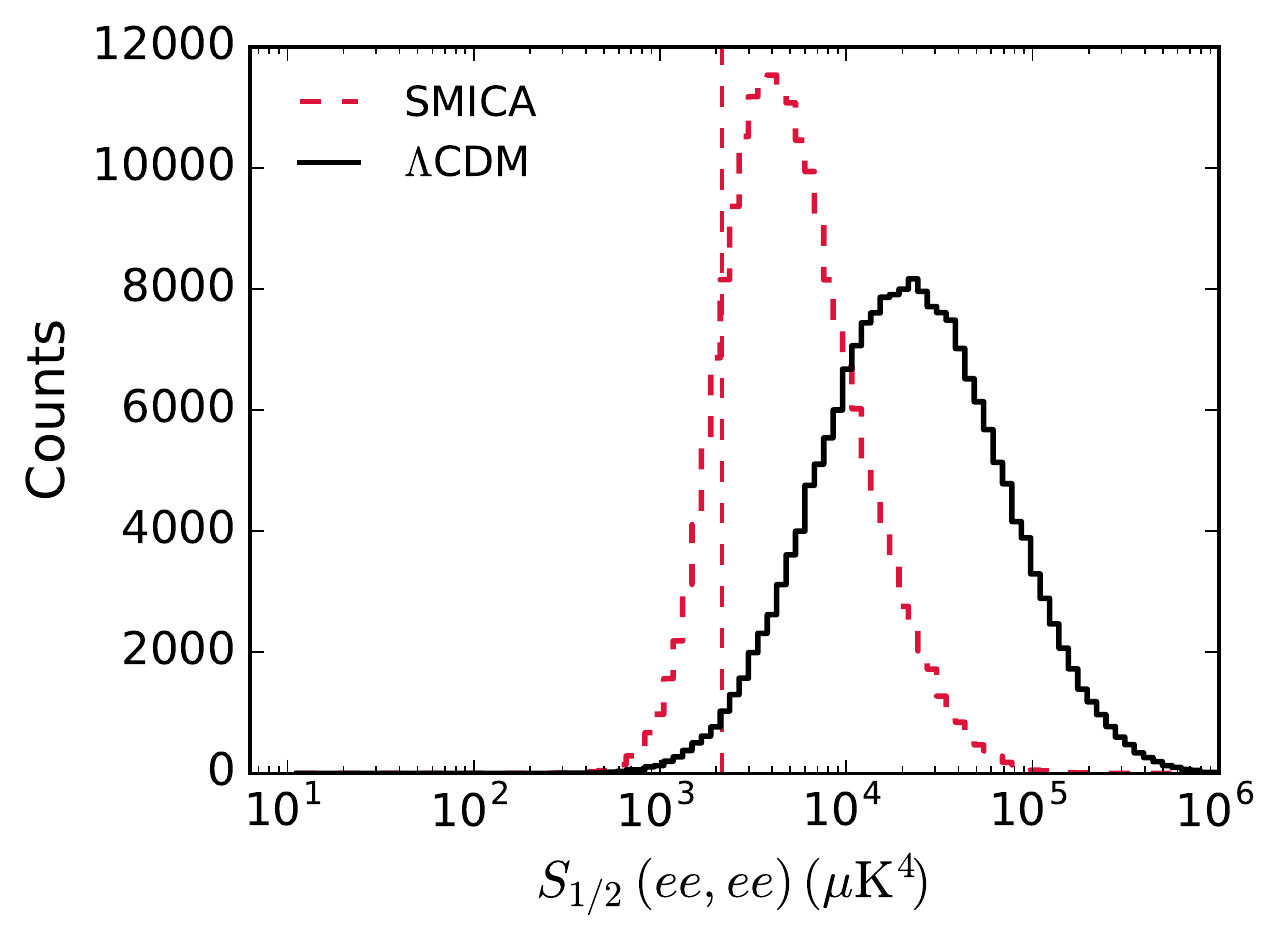}
  \includegraphics[width=\columnwidth]{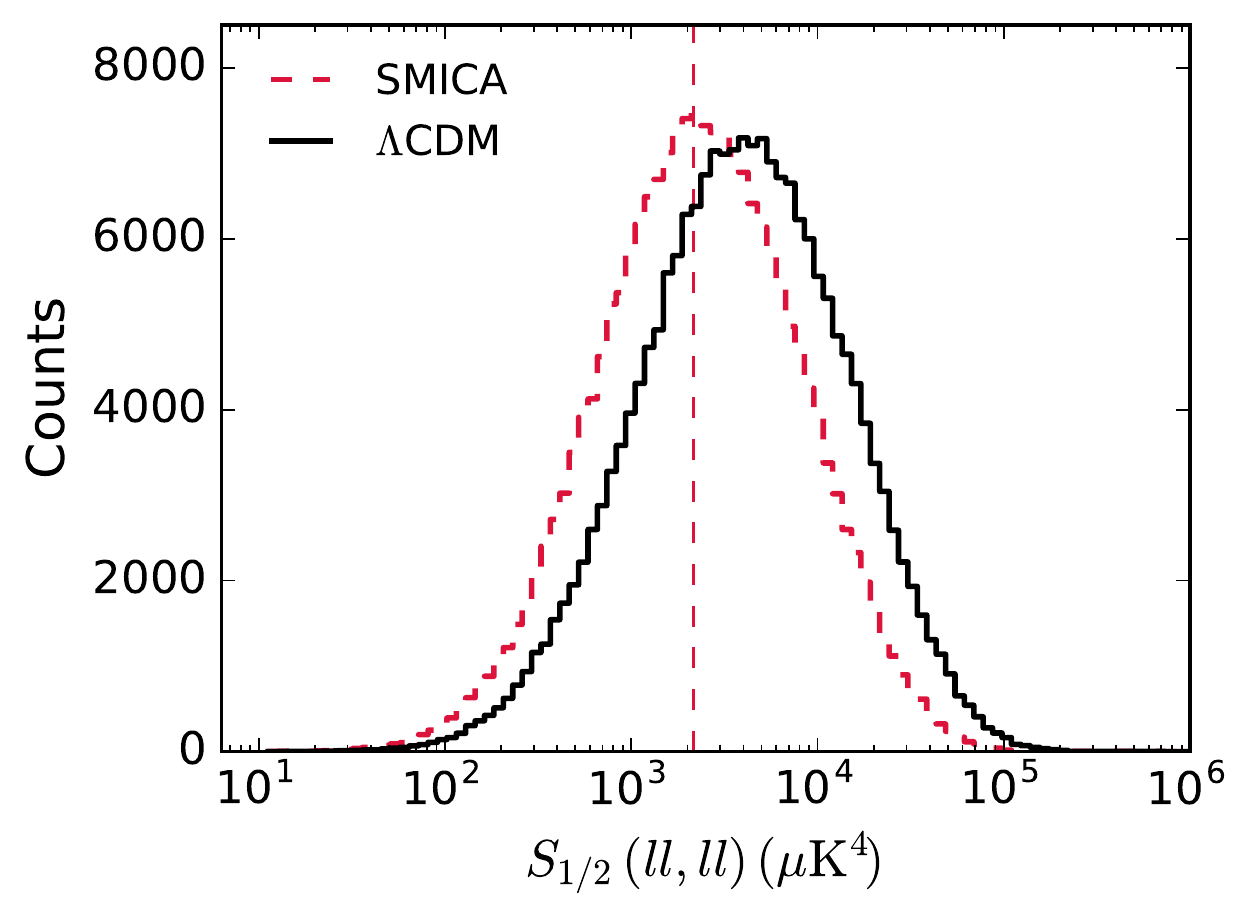}
  \includegraphics[width=\columnwidth]{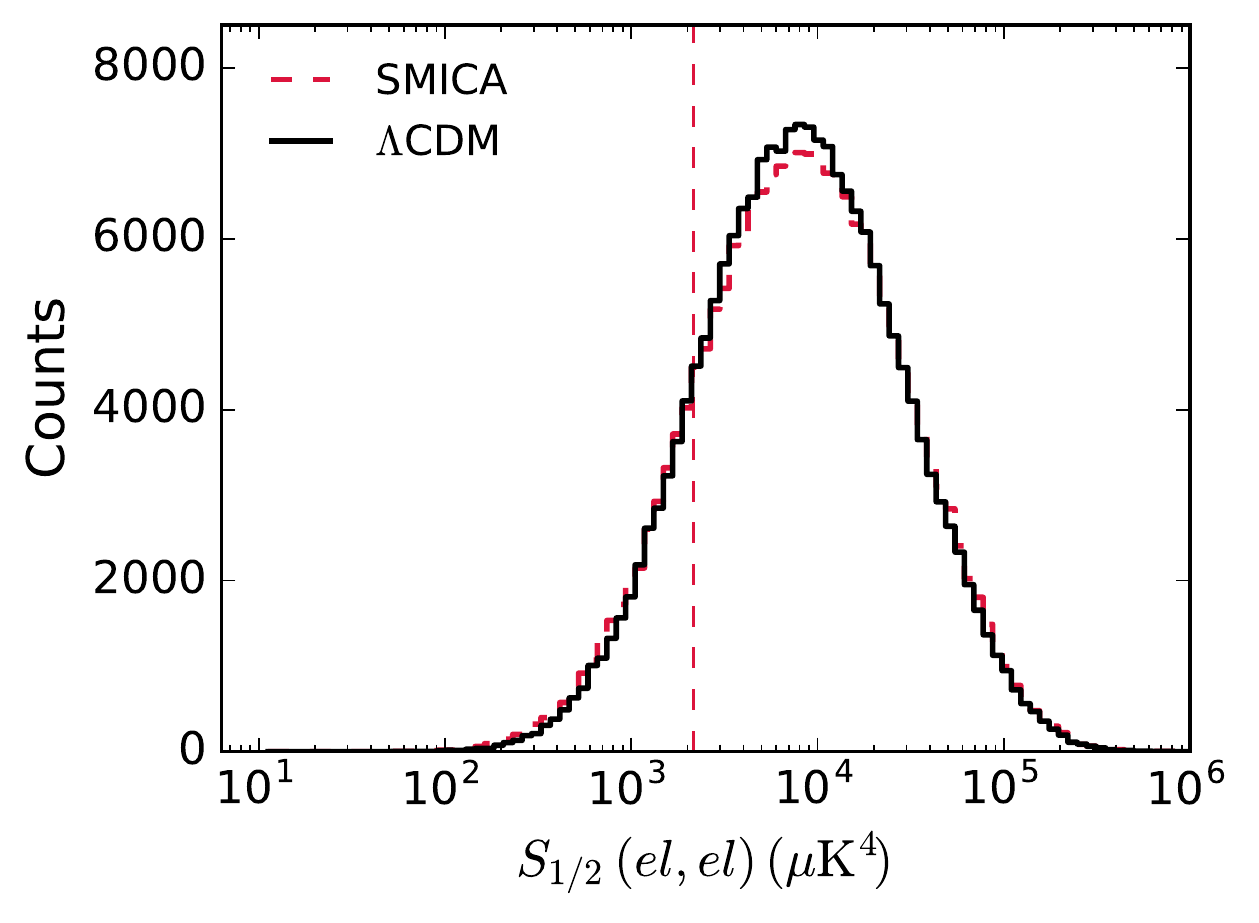}
    \caption{
    Distributions of masked-sky $\Shalf$ from $200\,000$
    \sI\ realizations  (solid, black lines) and \sII\ realizations (dashed,
    red lines). 
    (For detailed descriptions of the realizations, see section \ref{sec:Realizations}.) 
    The vertical lines represent the observed value of the masked-sky
    $\Shalf$ from the \smica\ map.
    Top left: The total $\Shalf$ probability distribution function (PDF)
    which includes contributions from SW, \eISW, and \lISW.  The
    \smica\ map has a single value so is only represented by the vertical
    line in this panel.
    Top right: The PDF of the auto-correlated, pure-early contribution,
    $\Shalf(\early\early,\early\early)$, which 
    shows a  suppression in the \sII\ realizations compared to the \sI\ ones.
    Bottom left: The PDF of $\Shalf(\late\late,\late\late)$ 
    which shows little change from the \sI\ realizations to the \sII\ ones -- 
    both distributions are consistent with the observed \smica\ value, 
    i.e. the observed $\Shalf$ value would not be anomalous in a \lISW-only sky. 
    Bottom right: The PDF of $\Shalf(\early\late,\early\late)$ 
    which is nearly identical in both the \sI\ realizations and the \sII ones.}
    \label{fig:S12}
\end{figure*}

\section{Results}

To begin, we first compare the \sI\ to the \sII\ realizations.  The
upper-left panel of Fig.~\ref{fig:S12} shows the $\Shalf$ probability
distribution function (PDF) for the \sI\ realizations.  Here the $\Shalf$
value for the \smica\ map with the \common\ mask, $\Shalf(\smica) = 2153
\muK^4$, is shown as the vertical dashed line and has a $p$-value of $0.58$
per cent, i.e. only that percentage of the \sI\ realizations had a smaller
$\Shalf$ than \smica.

The remaining panels in Fig.~\ref{fig:S12} display the auto-correlated
parts of $\Shalf$ for both the \sI\ and the \sII\ realizations.  The
biggest difference between the two is that
$\Shalf(\early\early,\early\early)$ is noticeably suppressed in the
\sII\ realizations compared to the \sI\ ones.  This is in part because
$\Shalf(\early\early,\early\early)$ is the largest single contribution to
$\Shalf$; however, note that in the \sII\ realizations, the mean of the PDF
of $\Shalf(\early\early,\early\early)$ is shifted \emph{below} that of
$\Shalf(\early\late,\early\late)$.  The latter is barely affected, and
perhaps even slightly \emph{increased} compared to the \sI\ realizations.
Meanwhile the $\Shalf(\late\late,\late\late)$ PDF is only slightly changed
and, more suggestively, if the \smica\ value (vertical line) were
pure $\Shalf(\late\late,\late\late)$, that value would be unremarkable.

These results taken together begin to suggest that the observed $\Shalf$
value is consistent with, and perhaps indicative of, a suppression of
early-time correlations.  If one could simply turn off early-time effects
in the theory without changing the late-time effect (e.g. somehow set
$\Cee^{\early\early}(\theta)=0$ with $\Cee^{\early\early}$ having the
obvious meaning of replacing $\Cl$ in (\ref{eqn:Ccorr}) with
$\Cl^{\early\early}$), so that the final $\Shalf$ distribution would be
completely $\Shalf(\late\late,\late\late)$, then the observed
\smica\ $\Shalf$ value would lie close to the mean of the PDF\@.

\begin{figure*}
  \includegraphics[width=\columnwidth]{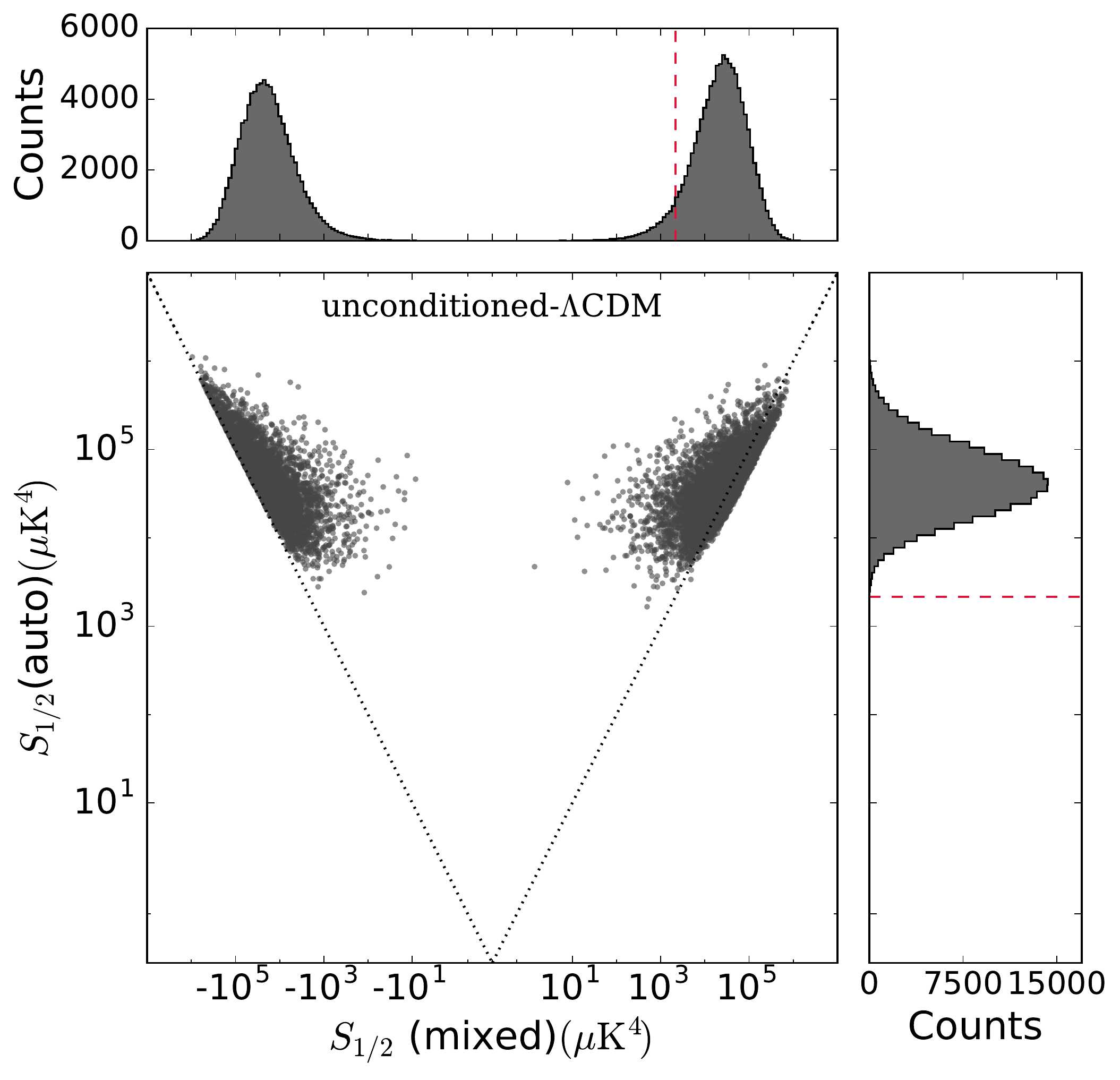}
  \includegraphics[width=\columnwidth]{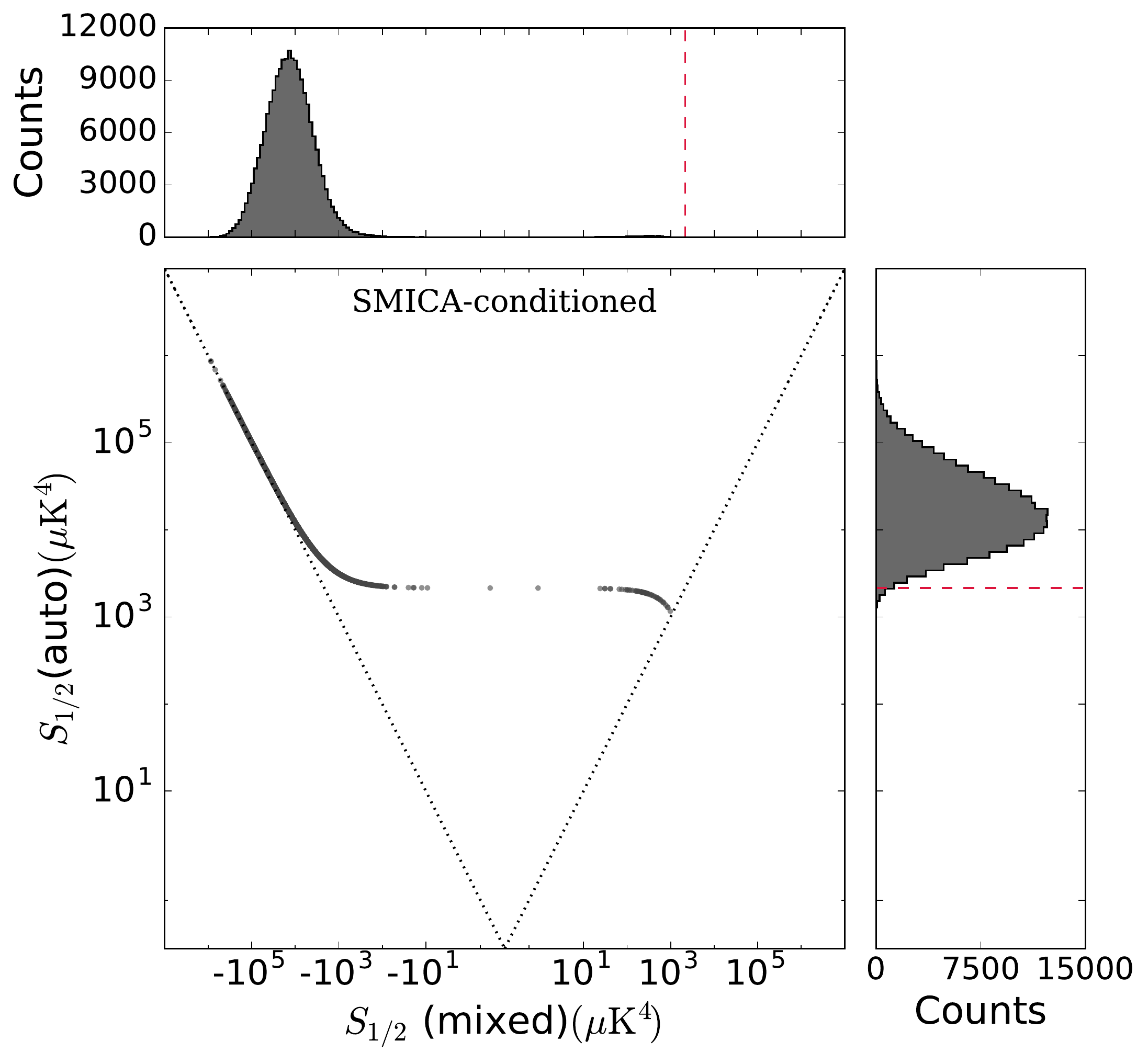}

    \caption{Scatter plots of masked-sky $\Shalf(\rmn{mixed})$
      \textit{versus} $\Shalf(\rmn{auto})$, as defined in
      Eqs.~(\ref{eq:S12-mixed}) and (\ref{eq:S12-auto}), for the
      \sI\ realizations (left) and for the \sII\ realizations (right).  For
      reference, the dotted lines show $\Shalf(\rmn{mixed})=\pm
      \Shalf(\rmn{auto})$, while red-dashed lines correspond to the
      observed masked-sky $\Shalf$ value from the \smica\ map.  For the
      \sI\ realizations we see roughly equal numbers of positive
      and negative values of $\Shalf(\rmn{mixed})$.  In contrast, for the
      \sII\ realizations there is an almost complete suppression of the
      positive values of $\Shalf(\rmn{mixed})$, and an almost perfect
      anti-correlation between $\Shalf(\rmn{mixed})$ and
      $\Shalf(\rmn{auto})$.  These two combine to yield the observed small
      value of $\Shalf$.  Note that \LCDM\ realizations constrained by a
      small $\Shalf$ sky (the \sIII\ realizations) consistently exhibit the
      same feature (not shown).}
    \label{fig:scatter}
\end{figure*}

To understand how the rare, small-$\Shalf$ realizations of \LCDM\ are
produced we consider the correlation between pairs of $\Shalf(X,Y)$.
Fig.~\ref{fig:scatter} shows a scatter-plot of the `mixed' parts with the
`auto' parts of $\Shalf$.  Here
\begin{equation}
  \Shalf(\rmn{mixed}) \equiv
  (\Shalf(\early\early,\late\late)+\Shalf(\early\early,\early\late)+
  \Shalf(\late\late,\early\late))
  \label{eq:S12-mixed}
\end{equation}
and
\begin{equation}
  \Shalf(\rmn{auto}) \equiv
  \Shalf(\early\early,\early\early)+\Shalf(\late\late,\late\late)+
  \Shalf(\early\late,\early\late).
  \label{eq:S12-auto}
\end{equation}
In the left panel, we see that in the \sI\ realizations there are roughly
equal numbers of positive and negative values of the mixed component, while
for the \sII\ ones (right panel) there is an almost complete suppression of
the positive values, supplemented by an almost perfect anti-correlation
between $\Shalf(\rmn{mixed})$ and $\Shalf(\rmn{auto})$.  Thus, the
combination of these two nearly cancel producing a small $\Shalf$.

The \sIII\ realizations allow us to determine whether the usual method of
producing a small $\Shalf$ found in the \sII\ realizations (the significant
cancellation) is typical in realizations of \LCDM\ with small $\Shalf$.
From the \sIII\ realizations, we found the same anti-correlation between
$\Shalf(\rmn{mixed})$ and $\Shalf(\rmn{auto})$ thus showing that if the
observed low $\Shalf$ is due to a fluke in \LCDM, it may at least be a
`typical fluke'.

\begin{figure}
  \includegraphics[width=\linewidth]{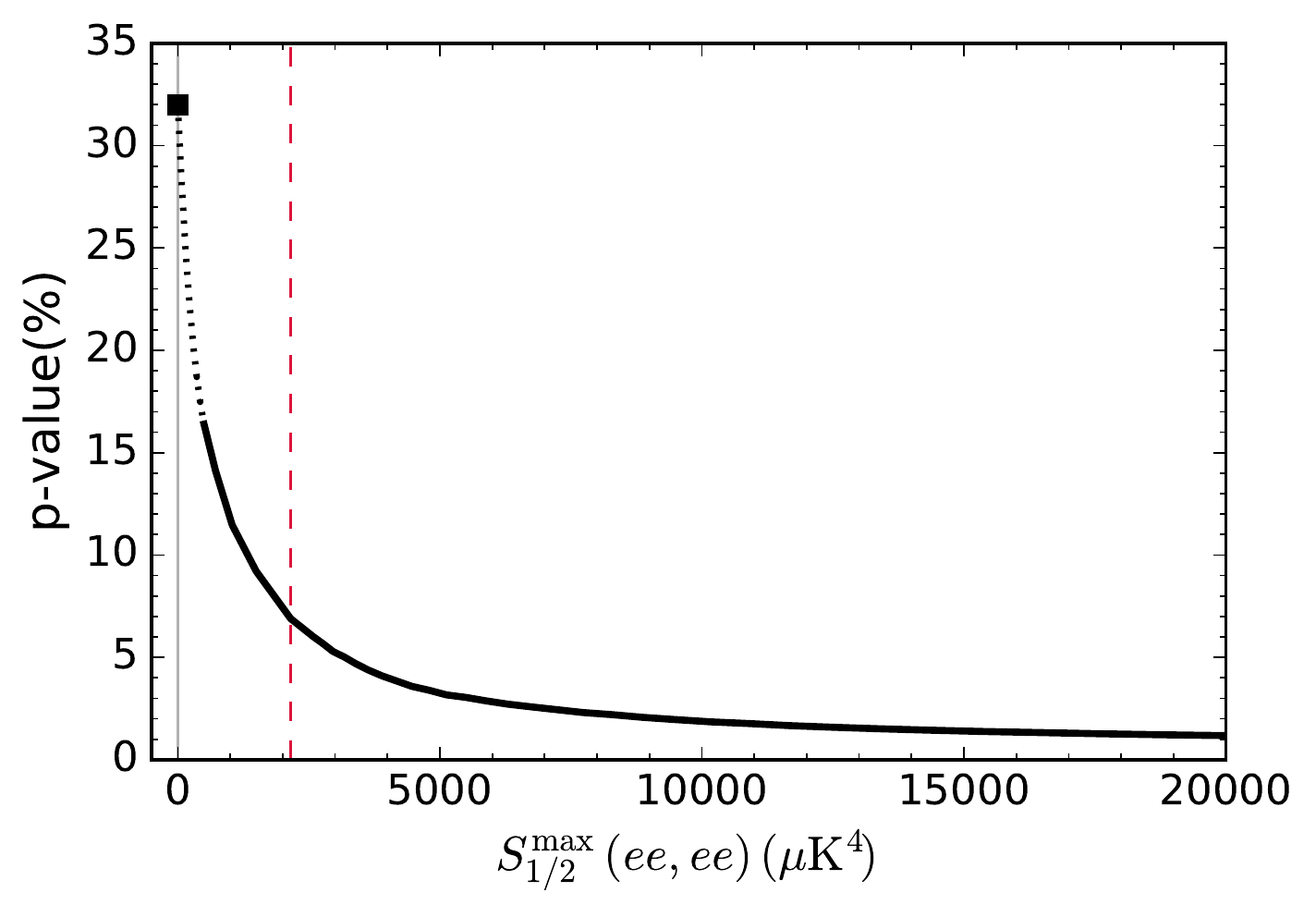}
  \caption{$p$-value of $\Shalf<\Shalf(\smica)=2153\muK^4$ among
    realizations of \LCDM\ with
    $\Shalf(\early\early,\early\early)<\Shalf^{\rmn{max}}(\early\early,\early\early)$
    as a function of that $\Shalf^{\rmn{max}}(\early\early,\early\early)$
    (solid line), derived from the \sIV\ realizations.  The small $p$-value
    of $\Shalf$ in the \sI\ realizations, $0.58$ per cent, is recovered at
    large $\Shalf^{\rmn{max}}(\early\early,\early\early)$, while an
    unremarkable $p$-value is found as
    $\Shalf^{\rmn{max}}(\early\early,\early\early)\to0$.  When
    $\Shalf^{\rmn{max}}(\early\early,\early\early)=0$, the $p$-value is
    that of a purely \lISW\ sky, ie.~approximately $32$ per cent.  The
    dotted curve is a suggestive interpolation between the $p$-values for
    $\Shalf^{\rmn{max}}(\early\early,\early\early)=0$ and
    $\Shalf^{\rmn{max}}(\early\early,\early\early)=500 \muK^4$.  The red
    dashed line marks $\Shalf$(\smica).  }
  \label{fig:pvaluevsearlyshalf}
\end{figure}

Finally, Fig.~\ref{fig:pvaluevsearlyshalf} shows the $p$-value of the
\smica\ $\Shalf$ value among the \sIV\ realization as a function of the
upper limit on $\Shalf(\early\early,\early\early)$.  A suggestive
interpolation between the lowest
$\Shalf^{\rmn{max}}(\early\early,\early\early)$ in the \sIV\ realizations
($500\muK^4$, the lowest value for which a sufficient number of
realizations was generated) and $\Shalf(\early\early,\early\early)=0$ (as
deduced from the $\Shalf(\late\late,\late\late)$ distribution from the
\sI\ realizations, see the bottom left panel of Fig.~\ref{fig:S12}) is
shown as a dotted curve.  We see that whereas the \smica\ value is rare
($p$-value of $0.58$ per cent) among generic \LCDM\ skies, it is reasonably
common ($p>10$ per cent) when $\Shalf(\early\early,\early\early)$ is
sufficiently small, approaching $p\approx 32$ per cent at
$\Shalf^{\rmn{max}}(\early\early,\early\early)=0$.  Because the
\Planck\ Release-2 \smica\ value of $\Shalf$ (with the \common\ mask) is
somewhat larger than that found in earlier maps, we have confirmed that the
same statement holds true even if we demand that $\Shalf<1600\muK^4$ (a
value more typical of \WMAP\ and \Planck\ Release-1 maps and masks).  This
strongly suggests that non-fluke explanations of the observed low values of
$\Shalf$ could reasonably expect to arise from new physics at or near the
LSS, without undue concern that the \lISW\ contribution would spoil the
explanation, though of course that would need to be checked in any specific
proposed model.

\section{Conclusions}

Using the \Planck\ Release-2 \smica, \commander, \sevem\ and
\nilc\ synthetic full-sky maps, and the best-fitting \Planck\ (TT + lowP +
lensing) parameters \citep{2015arXiv150201589P}, we have created sets of
realizations of the CMB temperature with the contributions from early-time
and late-time physics separated. The early-time physics is represented
through the Sachs-Wolfe and early-time integrated Sachs-Wolfe effects,
whereas the late-time physics is represented through the late-time
integrated Sachs-Wolfe effect.  For each generated sky we have calculated,
using the \Planck\ Release-2 \common\ (UT78) mask, the angular
auto-correlation functions for the early-time and late-time contributions,
$\Ccorr^{\early\early}(\theta)$ and $\Ccorr^{\late\late}(\theta)$, and the
cross-correlation function $\Ccorr^{\early\late}(\theta)$.  These have then
been combined into the six contributions to $\Shalf$, the statistic that
characterizes large-angle correlations of the CMB temperature, giving the
auto-correlation contributions $\Shalf(\early\early,\early\early)$,
$\Shalf(\early\late,\early\late)$, and $\Shalf(\late\late,\late\late)$, and
the mixed-correlation contributions $\Shalf(\early\early,\early\late)$,
$\Shalf(\early\early,\late\late)$, and $\Shalf(\early\late,\late\late)$.
Finally, for these $\Shalf(X,Y)$ we have produced probability distribution
functions (PDFs)\@.

Examining these PDFs we find that in \LCDM\ low values of cut-sky $\Shalf$,
such as are inferred from the \Planck\ Release-2 full-sky maps, typically
occur when, by chance, $\Shalf(\early\early,\early\early)$ is somewhat
suppressed, and the remaining auto-correlations are cancelled by the
mixed-correlations.  In other words, \LCDM\ realizations with low $\Shalf$
typically \emph{do not have} very low $\Shalf(ee,ee)$, instead, they also
include chance cancellations. The late-time contribution,
$\Shalf(\late\late,\late\late)$, need not be suppressed, suggesting that
there is no reason to expect unusually low amplitudes of late-time ISW,
given the low observed value of $\Shalf$.

Alternatively, when we examine \LCDM\ realizations in which
$\Shalf(\early\early,\early\early)$ is {\em constrained} to be suppressed
well below the value of $\Shalf$ inferred from the \Planck\ Release-2
full-sky maps, we find that the inferred \Planck\ value of $\Shalf$ is no
longer anomalous.  This suggests that physical explanations that operate to
suppress correlations on the last scattering surface (i.e. at high
redshifts), may be successful, in that they need not be
spoiled by the \lISW\ effect.  There is thus no obvious need to
(anti-)correlate early and late-time physics provided that the late-time
physics remains similar to that in the \LCDM\ model.

\section*{Acknowledgements}
GDS and MO'D are partially supported by Department of Energy grant
DE-SC0009946 to the particle astrophysics theory group at CWRU.  MO'D is
partially supported by the CAPES Foundation of the Ministry of Education of
Brazil.  Some of the results in this paper have been derived using the
\healpix\ \citep{2005ApJ...622..759G} package.

%%%%%%%%%%%%%%%%%%%%%%%%%%%%%%%%%%%%%%%%%%%%%%%%%%

%%%%%%%%%%%%%%%%%%%% REFERENCES %%%%%%%%%%%%%%%%%%

% The best way to enter references is to use BibTeX:

\bibliographystyle{mnras}%{mn2e_new}
\bibliography{s12_late_time_isw_final} % if your bibtex file is called example.bib

% Alternatively you could enter them by hand, like this:
% This method is tedious and prone to error if you have lots of references
%\begin{thebibliography}{99}
%\bibitem[\protect\citeauthoryear{Author}{2012}]{Author2012}
%Author A.~N., 2013, Journal of Improbable Astronomy, 1, 1
%\bibitem[\protect\citeauthoryear{Others}{2013}]{Others2013}
%Others S., 2012, Journal of Interesting Stuff, 17, 198
%\end{thebibliography}

%%%%%%%%%%%%%%%%%%%%%%%%%%%%%%%%%%%%%%%%%%%%%%%%%%

%%%%%%%%%%%%%%%%% APPENDICES %%%%%%%%%%%%%%%%%%%%%
\appendix
\section{Constrained Realizations}
\label{ap:alm}

Generating constrained, correlated Gaussian random variables is a
well-known topic.  Here we apply the standard approach to generating pairs
of early-time and late-time skies from a given total sky.  This discussion
is based closely on Appendix A of \cite{2013MNRAS.434.3590C}.

In \LCDM, the temperature anisotropies from early-time and late-time
physics, $\alm^\early$ and $\alm^\late$, respectively, 
are correlated and related to the associated full-sky anisotropies,
$\alm=\alm^\early+\alm^\late$.  Thus, given the full-sky $\alm$ and the
power spectra, $\Cl$, $\Cl^{\late\late}$, and $\Cl^{\early\late}$, the
correlated pairs, $\alm^\early$ and $\alm^\late$, can be generated.

Though typically written in terms of a complex basis, so that the $\alm$
are complex coefficients, it is more convenient to work in a real basis when
generating realizations.  Let $a_j$ represent a (real) coefficient in the
real basis. The index $j$ refers to the pair of standard spherical-harmonic
indices $(\ell,m)$, and takes values $0$ to $2\ell$, for each $\ell$. The
complex coefficients are constructed from the real coefficients as
\begin{equation}
  \alm = \left\{ \begin{array}{cl}
    a_0, & m = 0 \\
    \frac1{\sqrt{2}} \left( a_{2m-1} + \iimag\, a_{2m} \right), & m>0
  \end{array} \right. .
\end{equation}

Unconstrained realizations of \LCDM\ can then be generated in the real
basis as
\begin{eqnarray}
  a_j & = & \sqrt{\Cl} \zeta_1,
  \label{eq:ajF} \\
  a_j^{\late} & = & \frac{\Cl^{\early\late} + \Cl^{\late\late}}{\sqrt{\Cl}} \zeta_1 +
  \sqrt{\Cl^{\late\late}-\frac{(\Cl^{\early\late} + \Cl^{\late\late})^2}{\Cl}} \zeta_2,
  \label{eq:ajL}
\end{eqnarray}
where $\zeta_1$ and $\zeta_2$ are independent Gaussian random variables 
each drawn from a distribution with zero mean and unit variance.

However, once the full-sky $\alm$ are determined from observations the
$\alm^\late$ (or any other correlated quantity) is partially constrained.
Given particular values for the $\alm$ we need to generate realizations of
$\alm^{\late}$ that are consistent with these inputs.  In other words,
instead of randomly drawing a number for $\zeta_1$ when generating
$\alm^{\late}$, we instead \emph{solve} for $\zeta_1$ from
Eq.~(\ref{eq:ajF}).  This amounts to rewriting Eq. (\ref{eq:ajL}) as
\begin{equation}
 a_j^{\late}  =  \frac{\Cl^{\early\late} + \Cl^{\late\late}}{\Cl}a_j +
  \sqrt{\Cl^{\late\late}-\frac{(\Cl^{\early\late} + \Cl^{\late\late})^2}{\Cl}} \zeta_2,
\end{equation}
where $\zeta_2$ is still to be a random draw from a normal
distribution. Finally, for each draw, the early sky coefficients are
obtained by the difference $\alm^{\early} = \alm - \alm^{\late}$.

%%%%%%%%%%%%%%%%%%%%%%%%%%%%%%%%%%%%%%%%%%%%%%%%%%

% Don't change these lines
\bsp	% typesetting comment
\label{lastpage}
\end{document}